\documentclass[conference]{IEEEtran}
\usepackage[pdftex,dvipsnames]{xcolor}
\usepackage{todonotes}
\usepackage{verbatim}
\usepackage[htt]{hyphenat}
\usepackage[capitalize]{cleveref}
\usepackage{tabulary}
\usepackage{enumitem}

\usepackage{lipsum}
\usepackage{graphicx}
\usepackage{float}

\ifCLASSINFOpdf
\else
\fi
\begin{document}
%
\title{DANTE: Predicting Insider Threat using \\ LSTM on system logs}

\author{

\IEEEauthorblockN{Qicheng Ma}
\IEEEauthorblockA{The Rensselaer IDEA\\
Rensselaer Polytechnic Institute\\
New York, USA\\
maq3@rpi.edu}
\and
\IEEEauthorblockN{Nidhi Rastogi}
\IEEEauthorblockA{The Rensselaer IDEA\\
Rensselaer Polytechnic Institute\\
New York, USA\\
raston2@rpi.edu}
}


%


\maketitle

\begin{abstract}
Insider threat is one of the most pernicious threat vectors to information and communication technologies (ICT) across the world due to the elevated level of trust and access that an insider is afforded. This type of threat can stem from both malicious users with a motive as well as negligent users who inadvertently reveal details about trade secrets, company information, or even access information to malignant players. In this paper, we propose a novel approach that uses system logs to detect insider behavior using a special recurrent neural network (RNN) model. Ground truth is established using DANTE and used as baseline for identifying anomalous behavior. For this, system logs are modeled as a natural language sequence and patterns are extracted from these sequences. We create workflows of sequences of actions that follow a natural language logic and control flow. These flows are assigned various categories of behaviors - malignant or benign. Any deviation from these sequences indicates the presence of a threat. We further classify threats into one of the five categories provided in the CERT insider threat dataset. Through experimental evaluation, we show that the proposed model can achieve 93\% prediction accuracy.
\end{abstract}


%
\IEEEpeerreviewmaketitle

\section{Introduction}
Insider threat continues to top the security threat list year after year. The federal agency has increased its insider threat related spending and are on target to spend almost \$1 billion in fiscal year 2020 \cite{criste_2020}. When malicious activities are performed by insiders such as employees, contractors, temporary hires, interns, ex-employees, no amount of firewalls or multi-factor authentication can protect against their actions. An insider threat already has access to internal organizational information, has the motive as well as the means to sabotage the system for personal or professional gains. And the impact by an insider threat can be far more detrimental than any other threat vector. For example, they can compromise classified information, loss of revenues in the order of billions of dollars over a period of time, cause irreversible damage to reputation and many other forms of intangible damages. Covid-19 has further exacerbated the situation. Due to the remote nature of employee work environment, virtual private networks (VPNs) have replaced LANs, and so have the security posture around the employee devices. The speed at which the pandemic has spread, IT security administrators could not succeed in planning for this change and its subsequent impact on the security of the organization in advance. It has therefore, become easier to allow social engineering techniques to imposter an employee and gain access to internal company resources without raising an alarm.
\par
System and network logs actively record changes as well as the current state of IT systems and servers. Logs assist system administrators in monitoring activities and events, planning for outages and performance issues, and detecting abnormalities in the network. Since the information that can be mined from system and network logs is detailed and follows a time sequence, they can reveal information that points towards the possibility of intrusions, abnormal activities, and data theft. Therefore we utilize system logs for this reason. Another reason behind using system logs is that the process of log generation follows rules, logic, and control flows like a structured form of natural language. For example, an unsuccessful user login from a remote device to the database server will be recorded by the access control server as well as the database server. Therefore, this research uses system logs to present a novel anomaly detection-based approach to identify malicious insider behavior. Since log data is available in every networked organization and used to detect other intrusion activities, no additional data extraction is required to implement the proposed model, which is an additional advantage. \textsf{DANTE} uses a natural language-based model \cite{du2017deeplog} that uses system and network logs to generate sequences and create rule-based patterns. It captures sequences of activities across various log data to create workflow models for each separate task that can later be used to identify anomalous patterns and then be classified into one of the five categories of insider threats. 
\par
The \textsf{DANTE} deep neural network models these workflows using Long Short-Term Memory (LSTM) \cite{hochreiter1997long} and can learn long-range state over the sequence of information captured. It can then learn anomalous patterns and generate alert when deviations are detected. Experiments \textsf{DANTE} shows a 93\% accuracy in detecting insider behavior across all malicious categories. At this time, the model can successfully detect only known types of anomalies from the log entries. 
\section{Related Work}
Anomaly detection is frequently used to frame an insider threat detection problem due to the nature of the dataset - it is usually skewed and therefore captures more benign user behavior than malignant. Any outlier from a threshold indicates unacceptable behavior, which can be further investigated to determine the category of the threat as well as the root cause of the threat. Neural Networks have been used to classify and predict the presence of an Insider Threat \cite{tuor2017deep} as well as other cybersecurity threats \cite{veeramachaneni2016ai}.
\par
Recurrent Neural Networks (RNNs) \cite{mikolov2011extensions} allow persistence of information due to the presence of loops in their networks. This property connects the previous occurrence of information to the present one. An example of successful usage of RNNs is language models that predict the next word in a sentence, noted by keyboards in smartphones. Long Short Term Memory networks (LSTMs) \cite{hochreiter1997long} are a common type of artificial RNNs that can learn long-term information because of feedback connections in their networks. They differ from standard RNNs as they have four neural network layers instead of one in the repeating module.
\par
RNNs have been used to model different types of intrusions, such as real-time insider threat \cite{lu2019insider}, off-line intrusion detection \cite{staudemeyer2015applying}, etc. This is mainly due to the promising results from using LSTM models on this category of problems in the cybersecurity domain. Other than neural networks, in \cite{le2019machine}, researchers use supervised learning that incorporates feedback for insider threat detection.
\par
In \cite{tuor2017deep}, the authors use the LSTM approach for detecting insider threat alibi the dataset is structured. An interesting aspect of this research is that it assigns scores to each user's online session (sub-optimal approach), but requires an input of an expert to identify suspicious user sessions during training the dataset. Tabish et al. \cite{rashid2016new} use Hidden Markov Models (HMM) to learn 
benign user behavior. Any deviation from this implies an insider threat activity. While this requires fewer log entries to train the model, this approach requires an expert in the middle to differentiate normal activity from malicious. In addition to this, the variations in normal activity due to external factors have to be accounted for every time the model is trained. \cite{ma2020cybersecurity} use LSTM-CRF for named entity recognition from cybersecurity corpus, which remains a challenge largely due to unavailability of the structured and annotated security dataset. LSTM continues to be a favorite among cybersecurity researchers due to its promising results in detecting abnormalities and predicting the next event(s). \cite{khan2019scalable} have also used Convolutional-LSTM for detecting intrusion based threats in a network.
\par In this research, we continue to explore the advantage of using LSTM to identify network misuses using both global and user-specific latent threat vectors.
\section{Threat Model}
There are two principal components of \textsf{DANTE} - the RNN model and the data ingested in the model for classifying normal and abnormal behavior among users' system usage patterns. System and log data are generated by network devices, operating systems, applications distributed all across the network. While an insider can have access to one or more applications or computer devices, for this research, we assume that the logs are stored in a secure location that is not accessible and cannot be tampered with.
\par
System and log data are generated by various electronic devices, servers, and computers, which generate a sequence of log entries for individual users that are part of the system
generates a sequence of log entries for individual users that are part of the system. The assumption here is that the log data creation and storage are secure and protected from any adversarial attack. We also assume that despite the access levels provided to employees, where some of them own administrator-level access (system admin, their supervisor, IT managers) to servers and code that generated log files, as individuals they cannot modify the log files or the source code. The attack that we consider in this research is the insider threat who has access to various systems at different privilege levels.

\section{Data} \label{data}
Availability of a good dataset is crucial for monitoring tasks, and successful analysis. Likewise, Insider threat detection also benefits from the collection of a dataset that can enable predictions using machine learning models.
In line with this, we choose to use the CERT dataset \cite{collins2016common}, which is a decent size collection of synthetic insider threat test data sets. It contains both benign data from normal users as well as data from malicious actors \footnote{https://resources.sei.cmu.edu/library/asset-view.cfm?assetid=508099}. Also, for this research, we use this dataset since it contains a rich collection of log files from emails, HTTP connections, emails, device connections, user logon as well as psychometric tests (based on test results of OCEAN\footnote{https://en.wikipedia.org/wiki/Big\_Five\_personality\_traits}). 
\newline
The dataset had multiple releases (r1-r6), and generally speaking, the later releases included a super set of the earlier releases.
The insider threat detection methodology was proposed and evaluated on one of the six versions of this dataset. The dataset also includes the answer key file which comprises details of the malicious activity included in each dataset, including descriptions of the scenarios enacted and the identifiers of the synthetic users involved.
\newline
The r5.2 was chosen as the primary dataset for our research since, unlike the rest of the data sets, this one contained all attack scenarios described in the dataset. These are described below:
\begin{enumerate}
    \item Personnel worked after hours, used removable drive, and uploaded data to \texttt{wikileaks.org}. Left the organization soon after.
    \item Personnel visited job websites and solicited employment from a competitor(s). An abnormal increase in the usage of removable drive for data transfer. Left the organization soon after.
    \item Disgruntled system administrator, who downloaded a keylogger, used a removable drive for data transfer to their supervisor's computer. Collected key logs were used to log in as their supervisor. Alarming mass emails were sent out that caused panic in the organization. Left workplace right after this incident.
    \item Over a period of three months, personnel frequently logged into other user's computers. Searched and forwarded files on a personal email address.
    \item Personnel laid off. Uploaded documents to \texttt{Dropbox}.
\end{enumerate}

\subsection{About the Dataset}
Since we primarily use revision 5.2 of the dataset, the following description applies only to this version. We specifically employ the following data sources as input to our detection models - web history (\texttt{http.csv}), email log (\texttt{email.csv}), device access (\texttt{device.csv}), logon (\texttt{logon.csv}), file access (\texttt{file.csv}), and user data in the LDAP folder. While details of each of the log files can be gained from pulling the data from the link provided earlier, below we describe in detail some of the features that we captured before pre-processing:
\begin{enumerate}
   \item \texttt{LDAP} - employee name, user ID, email, role, Projects, business unit, functional unit, department, team, supervisor.
   \item \texttt{http.csv} - event ID (20 character long hash), time stamp, user ID, PC ID, web-page URL, activity ("WWW Visit", "WWW Download", "WWW Upload".), content.
    \item \texttt{email.csv} - event ID, time stamp, user ID, PC ID, to, cc, bcc, from, activity (Send, Receive or View), size, attachments, content.s
    \item \texttt{device.csv} - event ID, time stamp, user ID, PC ID, file-tree, activity (connect or disconnect)
    \item \texttt{logon.csv} - event ID, time stamp, user ID, PC ID, activity (logon or logoff activity).
    \item \texttt{file.csv} - event ID, time stamp, user ID, PC ID, filename, activity (open, copy, write, delete), to removable media, from removable media, content (hexadecimal encoded file header and a list of content keywords).
   
\end{enumerate}
\subsection{Data pre-processing}
The available data sources described in the previous section, while in a tabulated format, require processing to make them suitable for ingesting in our prediction model. For example, user-specific information across different sources was collated based on their relationship with other users (e.g. supervisor or not), logon or logoff into their personal computers and other computers, their designation (administrator with access to multiple computers), hours of activity, a sequence of activities across various logs, etc. For a given user - computer IDs that were accessed, domain category of the website accessed (described next), email domain category - internal or external to the organization, and so on. 
\par
In our model, we also identify acceptable working hours based on rules for both weekdays, weekends, and holidays. Websites were categorized based on their primary activity - job-hunting websites, hacktivist websites, and file-sharing websites. This allows normalization and quantification of data points which would otherwise be a challenge to utilize in modeling. Each user's activity is tracked by user activity, and not based on daily, weekly, or monthly activity. This allows detection at a much granular level, and also real-time detection of anomalous behavior when we prepare our model for real-time detection at a later stage.

\begin{figure*}
    \centering
    \frame{\includegraphics[width= .9\textwidth]{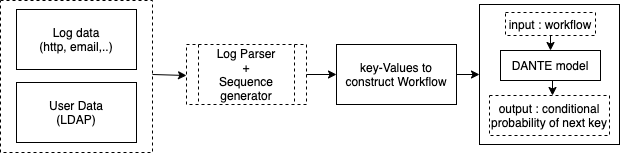}}
    \caption{Block diagram showing various \textsf{DANTE} components.}
    \label{fig:architecture}
\end{figure*}

\section{DANTE Architecture}
The \textsf{DANTE} architecture as can be seen in Figure~\ref{fig:architecture} shows the main components as well as the end-to-end implementation flow. 
\subsection{Log Parsing}
Log files can be distributed across various electronic devices in a networked system, it is important to collect them concurrently. Log entries are then parsed into structured representation and later filtered for each user who uses the system in a given time frame. The logs collected for each user should follow a time-series before ingesting them into the anomaly detection model. Prior work shows that from each log entry, a message type can be extracted as the log key \cite{du2016spell,lou2010mining}. When logs are generated from a source code, print statements are used to write statements that match an event, along with values that describe that event. For example, for a print code - \texttt{printf("Logged into \%f.”, PC\_ID")}, the value of PC\_ID is \textbf{\textit{t\textsubscript{n}}} and the statement \textit{"Logged into *."} is the key, \textbf{\textit{k}}. See Table \ref{table:QA} for more examples.

\begin{table}[hbt!]
\vspace{0.25in}
\centering
\begin{tabular}{{p{4cm}|p{1cm}|p{2.5cm}}}
\hline
  {\bf Log Message} & {\bf key} & \bf values\\
  \hline
  \textbf{\textit{t\textsubscript{1}} Logoff from \textcolor{blue}{PC-003}} & \textit{k}\textsubscript{1} & [t\textsubscript{1}, PC-003] \\ 
 \hline 
  \textbf{\textit{t\textsubscript{2}} Delete \textcolor{blue}{file-GdY68}} & \textit{k}\textsubscript{2} & [t\textsubscript{2}, file-GdY68] \\ 
 \hline
   \textbf{\textit{t\textsubscript{3}} Access \textcolor{blue}{file-HJd84}} & \textit{k}\textsubscript{3} & [t\textsubscript{3}, file-HJd84] \\ 
 \hline
\end{tabular}
\vspace{0.15in}
\caption{Exemplar entries from log files}
\label{table:QA}
\end{table}

\subsection{Model Description}
The architecture is shown in Figure \ref{fig:architecture} with three main components: log key-value generation from log data, log parsing, workflow construction, and anomaly detection using \textsf{DANTE}. It uses LSTM as the main neural network model to process sequences of logs and classify normal and outlier behavior in the logs. Since logs can sometimes have technical issues that might result in a delay in user event record collection or may encounter offline users (especially when they are a threat), it is ideal to use LSTMs in such scenarios. This results in premature convergence to poor solutions (called vanishing gradient problem) when training traditional RNNs. However, LSTMs were developed to deal with this issue as they are relatively insensitive to gap length.
\subsection{Workflows}
\begin{figure*}
    \centering
    \frame{\includegraphics[width= 1\textwidth]{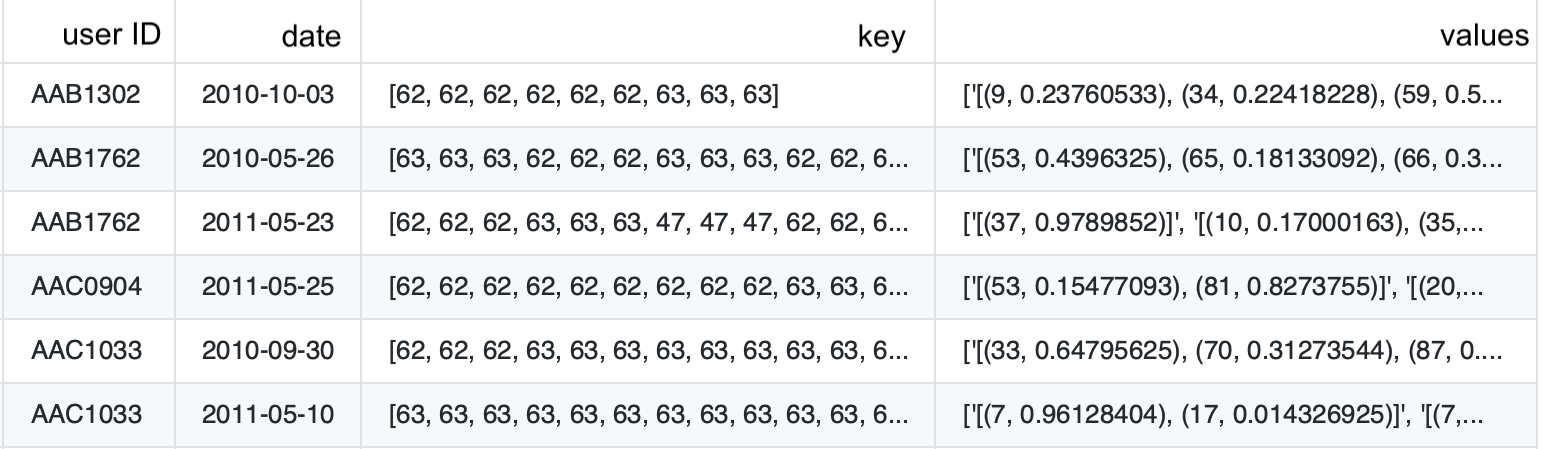}}
    \caption{A snippet of key value generation. For each user, in a given day, each key is associated with a value and the probability of it's occurrence.}
    \label{fig:key-value}
\end{figure*}

In this section, we describe the creation of log key sequences that form workflows of each user per day. Log printing source code has a finite number of print statements, and therefore, can print a finite number of log keys. The set  \textit{K = {\textit{k}\textsubscript{1},\textit{k}\textsubscript{2},...,\textit{k}\textsubscript{n}}} is that finite set of unique log keys. In Figure \ref{fig:key-value}, we show a few examples of how the key-value pairs are generated and used to create a workflow.

\subsection{Model Training}
The structuring of unstructured logged events requires pre-processing of the data by first collecting unique ids, emails, and "PC-IDs" on the network. Both active and dormant accounts are counted. The log events are then parsed using rule-based mining and assigned to specific users. While this approach is found to be effective in creating a sequence of activities for every single user, it assumes that the insider threat is not working in cahoots with other employees. With every users' temporal data assigned to them, log-key sequences are created and added to the existing training dataset. This sequence suggests, with high probability, an execution path or an execution order of print statements that write into various logs. LSTM is used to train for recent context as a multi-class classifier that uses keys and log-key values as input and the output is a probability distribution over log keys, \textit{k} that will be next in sequence. This approach is influenced by n-gram models that predict subsequent words when an initial set of text has been used for training~\cite{bickel2005predicting,manning1999foundations}. For \textsf{DANTE}, we extend the \textit{n-gram} model to create a sequence of log events and eventually classify and predict the presence of a threat. Here, each log key entry, \textit{k} corresponds to a word taken from the vocabulary \textit{K}. The \textit{n-gram} model works on the intuition that recent predecessors influence the prediction of next word (or sequence) for a given word, and therefore having a history of \textit{n} words suffices.
\subsection{LSTM}
An LSTM-based model can encode complex patterns, such as log sequences, and maintain long-range state over a sequence. Therefore, we use LSTM models to detect anomalous behavior in workflows created using log-key sequences. An input consists of a window \textit{w} of \textit{h} log keys, and an output is the log key value that comes right after \textit{w}. We use the categorical cross-entropy loss for training. After training is done, we can predict the output for an input \textit{(w = {m\textsubscript{t-h},...,m\textsubscript{t-1}})} using a layer of \textit{h} LSTM blocks. Each log key in \textit{w} feeds into a corresponding LSTM block in this layer.
\newline
\textit{Anomaly detection} - For anomaly detection from log keys, the input is a sequence of log keys of length \textit{h} from recent history, and the output is a probability distribution of all possible log key values. The training set from benign users forms the baseline. The difference between a prediction and an
observed parameter value vector is measured by the mean square
error (MSE). At deployment, if the error between a prediction and an observed value vector is within a high-level of the confidence interval of the above Gaussian distribution, the parameter value vector of the incoming log entry is considered normal and is considered abnormal otherwise.

\section{Experiments and Results}
We deployed \textsf{DANTE} model on google colab\footnote{https://colab.research.google.com} platform to execute experiments. The log entries are divided into three sets - one for training, one for testing, and one for validation. During testing, \textsf{DANTE} measures the gaussian distribution of MSE between the proposed key and the actual key in a given workflow. If it is within an acceptable confidence interval of the MSE, the next key is accepted, else considered anomalous.
\par
The GitHub code has been made available as open source\footnote{https://github.com/aiforsec/InsiderThreat} and can be downloaded and used in tandem with the CERT dataset r5.2.

\section{Results}

In Table \ref{table:Summary}, we summarize the number of users identified as insider threats and the number of logs classified as threat scenarios described in Section \ref{data}. Note, four out of five cases were classified.

\begin{table}[hbt!]
\vspace{0.1in}
\centering
\begin{tabular}{{p{1.5cm}|p{2cm}|p{2cm}}}
   \hline
  {\bf } & {\bf No. of Users} & {\bf No. of Logs} \\
  \hline
  {\bf Total  } & {2000 } & {79856704 } \\
  \hline
    {\bf Benign } & { 1901} & {79846376 } \\
  \hline
    {\bf S-1  } & {29 } & {486 } \\
  \hline  
  {\bf  S-2} & {30 } & {6477 } \\
  \hline  
  {\bf S-3 } & {10 } & {184 } \\
  \hline  
  {\bf S-4 } & {30} & { 3181}\\
 \hline
\end{tabular}
\vspace{0.15in}
\caption{Summary of predicted insiders, log entries for each scenario.}
\label{table:Summary}
\end{table}

The model shows an accuracy of 93.29\% with an average loss of 0.230025 for 500 epochs.

\begin{figure}[h!]
    \centering
    \includegraphics[width= .45\textwidth]{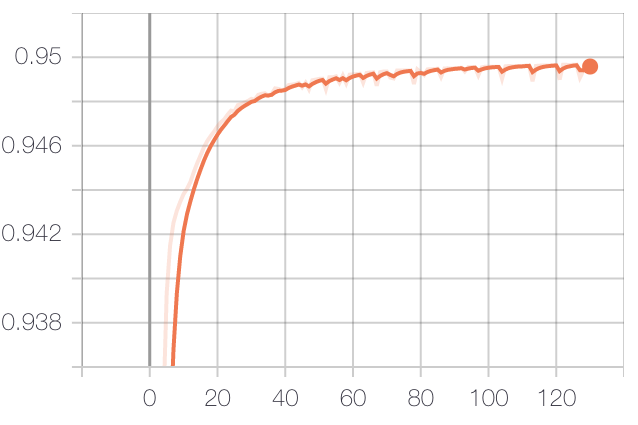}
    \caption{Evaluation of \textsf{DANTE} - Epoch vs Accuracy (approx. 93\%)}
    \label{fig:train-accuracy}
\end{figure}

Figure \ref{fig:train-accuracy} shows the detection accuracy of the model. Also, in Figure \ref{fig:train_avgLoss}, we show the average loss as the training dataset increases. The accuracy is approximately 93\% and the average loss is less than .23.

\begin{figure}[h]
    \centering
    \includegraphics[width= .45\textwidth]{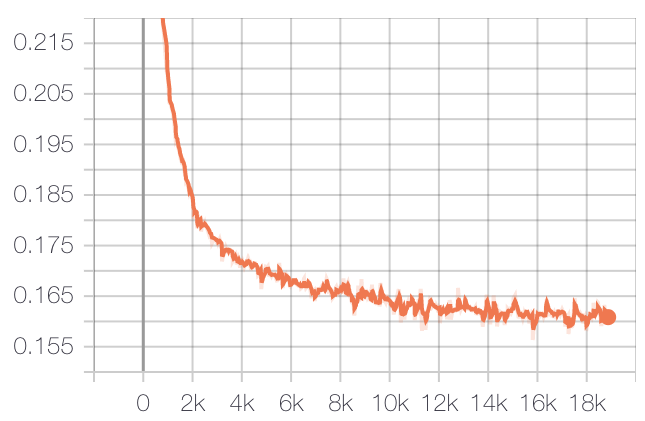}
    \caption{Evaluation of \textsf{DANTE} - Training dataset vs Average Loss (approx 0.25).}
    \label{fig:train_avgLoss}
\end{figure}

\begin{figure}[h]
    \centering
    \includegraphics[width= .4\textwidth]{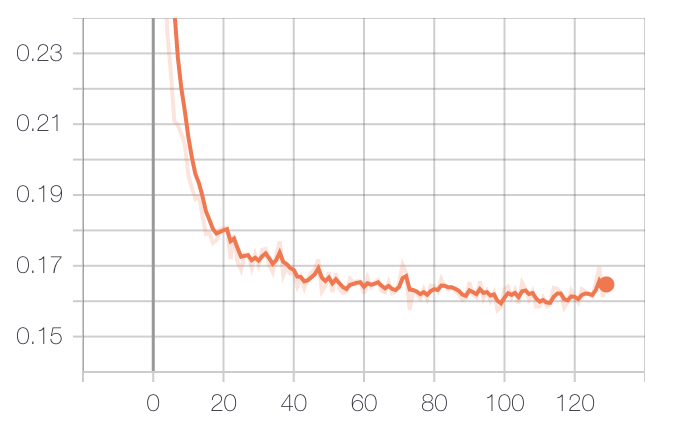}
    \caption{Evaluation of \textsf{DANTE} - Epoch count vs Average Loss.}
    \label{fig:train_avgLoss}
\end{figure}

\section{Discussion and Future Work}
This is a work in progress and is part of a larger research road map on using AI models for insider threat detection. In this paper, we present the initial, exploratory part of \textsf{DANTE} model where logs are used to create a sequence of events for each user for a given time frame (usually 24 hours). Since log creation follows control logic and flow, it is safe to assume that this flow can be dealt in the same way as natural language models are. However, there are several challenges that need to be addressed in the current version of \textsf{DANTE}. A shortcoming of anomaly detection, generally, is that the baseline has been fully modeled, and hence anything outside the threshold will be considered as a potential threat. Or the assumption is that the baseline model is continuously updated with new acceptable IT scenarios. This approach suffers from high false positives and therefore, false alarms. We will be exploring clustering techniques that identify various behaviors of the users without a-priori knowledge of benign user behavior. The second problem we are trying to solve is the problem of loops within key-values, where the same value might reoccur. A third problem we're looking into is the granularity of the log key-value pair. The current approach uses the smallest level of log sequence, and this might lead to overlooking new patterns by new users who join the organization with new roles and responsibilities, and hence a new work-flow. Some of these challenges include the assumption of unchanged source code that is used to generate log statements. Further, as mentioned earlier in the paper, using time-series event logs for individual users assumes that more than one user is not operating in cahoots. An insider knows no bounds, and with the right privileges, they can poison both the model as well as the data that is used to create the baseline (benign behavior) or models for various malignant behavior. So we would enhance the threat model with these scenarios. \textsf{DANTE} can be improved further with an online feedback mechanism that reduces the chances of false positives and false negatives.


%

\section*{Acknowledgment}
This work is supported by IBM Research AI through the AI Horizons Network research grant.
\bibliographystyle{IEEEtran}
\bibliography{references}

\begin{thebibliography}{10}
\providecommand{\url}[1]{#1}
\csname url@samestyle\endcsname
\providecommand{\newblock}{\relax}
\providecommand{\bibinfo}[2]{#2}
\providecommand{\BIBentrySTDinterwordspacing}{\spaceskip=0pt\relax}
\providecommand{\BIBentryALTinterwordstretchfactor}{4}
\providecommand{\BIBentryALTinterwordspacing}{\spaceskip=\fontdimen2\font plus
\BIBentryALTinterwordstretchfactor\fontdimen3\font minus
  \fontdimen4\font\relax}
\providecommand{\BIBforeignlanguage}[2]{{%
\expandafter\ifx\csname l@#1\endcsname\relax
\typeout{** WARNING: IEEEtran.bst: No hyphenation pattern has been}%
\typeout{** loaded for the language `#1'. Using the pattern for}%
\typeout{** the default language instead.}%
\else
\language=\csname l@#1\endcsname
\fi
#2}}
\providecommand{\BIBdecl}{\relax}
\BIBdecl

\bibitem{criste_2020}
\BIBentryALTinterwordspacing
L.~Criste, ``Insider threat market to top \$1 billion in fiscal 2020: This
  is,'' Aug 2020. [Online]. Available:
  \url{https://about.bgov.com/news/insider-threat-market-to-top-1-billion-in-fiscal-2020-this-is/}
\BIBentrySTDinterwordspacing

\bibitem{du2017deeplog}
M.~Du, F.~Li, G.~Zheng, and V.~Srikumar, ``Deeplog: Anomaly detection and
  diagnosis from system logs through deep learning,'' in \emph{Proceedings of
  the 2017 ACM SIGSAC Conference on Computer and Communications Security},
  2017, pp. 1285--1298.

\bibitem{hochreiter1997long}
S.~Hochreiter and J.~Schmidhuber, ``Long short-term memory,'' \emph{Neural
  computation}, vol.~9, no.~8, pp. 1735--1780, 1997.

\bibitem{tuor2017deep}
A.~Tuor, S.~Kaplan, B.~Hutchinson, N.~Nichols, and S.~Robinson, ``Deep learning
  for unsupervised insider threat detection in structured cybersecurity data
  streams,'' \emph{arXiv preprint arXiv:1710.00811}, 2017.

\bibitem{veeramachaneni2016ai}
K.~Veeramachaneni, I.~Arnaldo, V.~Korrapati, C.~Bassias, and K.~Li, ``Ai\^{} 2:
  training a big data machine to defend,'' in \emph{2016 IEEE 2nd International
  Conference on Big Data Security on Cloud (BigDataSecurity), IEEE
  International Conference on High Performance and Smart Computing (HPSC), and
  IEEE International Conference on Intelligent Data and Security (IDS)}.\hskip
  1em plus 0.5em minus 0.4em\relax IEEE, 2016, pp. 49--54.

\bibitem{mikolov2011extensions}
T.~Mikolov, S.~Kombrink, L.~Burget, J.~{\v{C}}ernock{\`y}, and S.~Khudanpur,
  ``Extensions of recurrent neural network language model,'' in \emph{2011 IEEE
  international conference on acoustics, speech and signal processing
  (ICASSP)}.\hskip 1em plus 0.5em minus 0.4em\relax IEEE, 2011, pp. 5528--5531.

\bibitem{lu2019insider}
J.~Lu and R.~K. Wong, ``Insider threat detection with long short-term memory,''
  in \emph{Proceedings of the Australasian Computer Science Week
  Multiconference}, 2019, pp. 1--10.

\bibitem{staudemeyer2015applying}
R.~C. Staudemeyer, ``Applying long short-term memory recurrent neural networks
  to intrusion detection,'' \emph{South African Computer Journal}, vol.~56,
  no.~1, pp. 136--154, 2015.

\bibitem{le2019machine}
D.~C. Le and A.~N. Zincir-Heywood, ``Machine learning based insider threat
  modelling and detection,'' in \emph{2019 IFIP/IEEE Symposium on Integrated
  Network and Service Management (IM)}.\hskip 1em plus 0.5em minus 0.4em\relax
  IEEE, 2019, pp. 1--6.

\bibitem{rashid2016new}
T.~Rashid, I.~Agrafiotis, and J.~R. Nurse, ``A new take on detecting insider
  threats: exploring the use of hidden markov models,'' in \emph{Proceedings of
  the 8th ACM CCS international workshop on managing insider security threats},
  2016, pp. 47--56.

\bibitem{ma2020cybersecurity}
P.~Ma, B.~Jiang, Z.~Lu, N.~Li, and Z.~Jiang, ``Cybersecurity named entity
  recognition using bidirectional long short-term memory with conditional
  random fields,'' \emph{Tsinghua Science and Technology}, vol.~26, no.~3, pp.
  259--265, 2020.

\bibitem{khan2019scalable}
M.~A. Khan, M.~Karim, Y.~Kim \emph{et~al.}, ``A scalable and hybrid intrusion
  detection system based on the convolutional-lstm network,'' \emph{Symmetry},
  vol.~11, no.~4, p. 583, 2019.

\bibitem{collins2016common}
M.~Collins, M.~Theis, R.~Trzeciak, J.~Strozer, and J.~Clark, ``Common sense
  guide to mitigating insider threats fifth edition,'' 2016.

\bibitem{du2016spell}
M.~Du and F.~Li, ``Spell: Streaming parsing of system event logs,'' in
  \emph{2016 IEEE 16th International Conference on Data Mining (ICDM)}.\hskip
  1em plus 0.5em minus 0.4em\relax IEEE, 2016, pp. 859--864.

\bibitem{lou2010mining}
J.-G. Lou, Q.~Fu, S.~Yang, Y.~Xu, and J.~Li, ``Mining invariants from console
  logs for system problem detection.'' in \emph{USENIX Annual Technical
  Conference}, 2010, pp. 1--14.

\bibitem{bickel2005predicting}
S.~Bickel, P.~Haider, and T.~Scheffer, ``Predicting sentences using n-gram
  language models,'' in \emph{Proceedings of Human Language Technology
  Conference and Conference on Empirical Methods in Natural Language
  Processing}, 2005, pp. 193--200.

\bibitem{manning1999foundations}
C.~Manning and H.~Schutze, \emph{Foundations of statistical natural language
  processing}.\hskip 1em plus 0.5em minus 0.4em\relax MIT press, 1999.

\end{thebibliography}
\end{document}